# Class-B cable-driving SQUID amplifier

**Mikko Kiviranta**
VTT, Tietotie 3, 02150 Espoo, Finland

E-mail: Mikko.Kiviranta@vtt.fi

**Abstract.** We suggest a SQUID amplifier configuration with improved power efficiency, for applications where cooling budgets are limited, in particular superconducting detector array readouts in space observatories. The suggested two-SQUID configuration keeps one SQUID in a zero-voltage state during a half-cycle of the input while the other SQUID generates voltage. For the other half-cycle the roles of the SQUIDs are reversed. The circuit is an electrostatic dual of a transistor-based amplifier operating in class-B. A proof-of principle demonstration at $T = 4.2$ K is presented.

## 1. Introduction

Some applications of SQUIDs involve signals with very large dynamic range together with a large bandwidth. Of these, multiplexed arrays of cryogenic detectors [1] located in space observatories such as ATHENA [2] are typical, and are further constrained by the amount of thermal power that can be dissipated. On orbit, the generated heat must be radiated to the space, and increase in heat generation rapidly escalates the mission launch mass and the mission cost. The SQUID amplifiers are almost always operated in class-A [3], in which the SQUID bias is chosen such that in absence of signal the SQUID dwells roughly at middle of its output voltage range, with significant static power dissipation. For the two half-cycles of the input flux signal, assumed sinusoidal, the SQUID voltage changes to positive and negative direction from the chosen setpoint, respectively, but remains finite over the whole cycle.

We have suggested in the past [4] a two-SQUID configuration where one of the SQUIDs is in the voltage state during a half-cycle of the signal, while the other SQUID remains superconductive. During the other half-cycle the roles of the SQUIDs are exchanged. This mode of operation is effectively an electrostatic dual of the class-B operation of transistor and vacuum tube amplifiers. An inactive SQUID is a perfect conductor while an inactive transistor is an almost perfect insulator, neither device dissipating power.

Our earlier circuit [4] was intended for a current-input amplifier as the subsequent stage, a typical configuration of a millikelvin stage of a two-stage SQUID tandem [5]. Here we demonstrate experimentally a configuration for a voltage-input subsequent amplifier, a typical configuration for the final SQUID stage of a multistage amplifier, which drives the cable from the cryostat to the room temperature low noise amplifier (LNA). Our aim is to improve power efficiency of the SQUID amplifier, i.e. the ratio of the available output signal power to the dissipated power.

## 2. Operating principle

The simplified schematic and characteristics of the most commonly used class-A SQUID configuration are presented in figs. 1a, 1d and 1g. In the actual experiments a 4-parallel 40-series SQUID array was used [6], which is equipped with a middle tap dividing the series array into a pair of 20-series sub-arrays. The VTT designation for the SQUID chip type is 'N2'. The middle tap is optional, and removal of its ground connection leads to a fully differential version of the standard SQUID readout scheme [7]. The $I_B \approx$ 100 µA bias current is in our experiment provided through 10 kΩ series resistors, by a biasing voltage and an inverted copy of the voltage via an inverting buffer. A disadvantage of the configuration is that the average voltage across the LNA inputs is not centered at zero, consuming the LNA dynamic range margin [8].

Another class-A circuit for SQUIDs [9], [10] is shown in figs. 1b, 1e and 1h, a SQUID version of the traditional push-push valve circuit [11]. Both sub-arrays are biased with the same polarity of bias current, but the input coil polarity is reversed for the second sub-array. Positive change in the applied flux hence causes the voltage of the first sub-array to increase and the second sub-array to decrease. The resulting differential voltage is centered at zero at the LNA input, and the LNA dynamic range is used efficiently [8]. We also note that the inherent non-linearity in the flux response of a current-biased dc SQUID gets cancelled in the first order [12].

A class-B cable-driving SQUID amplifier

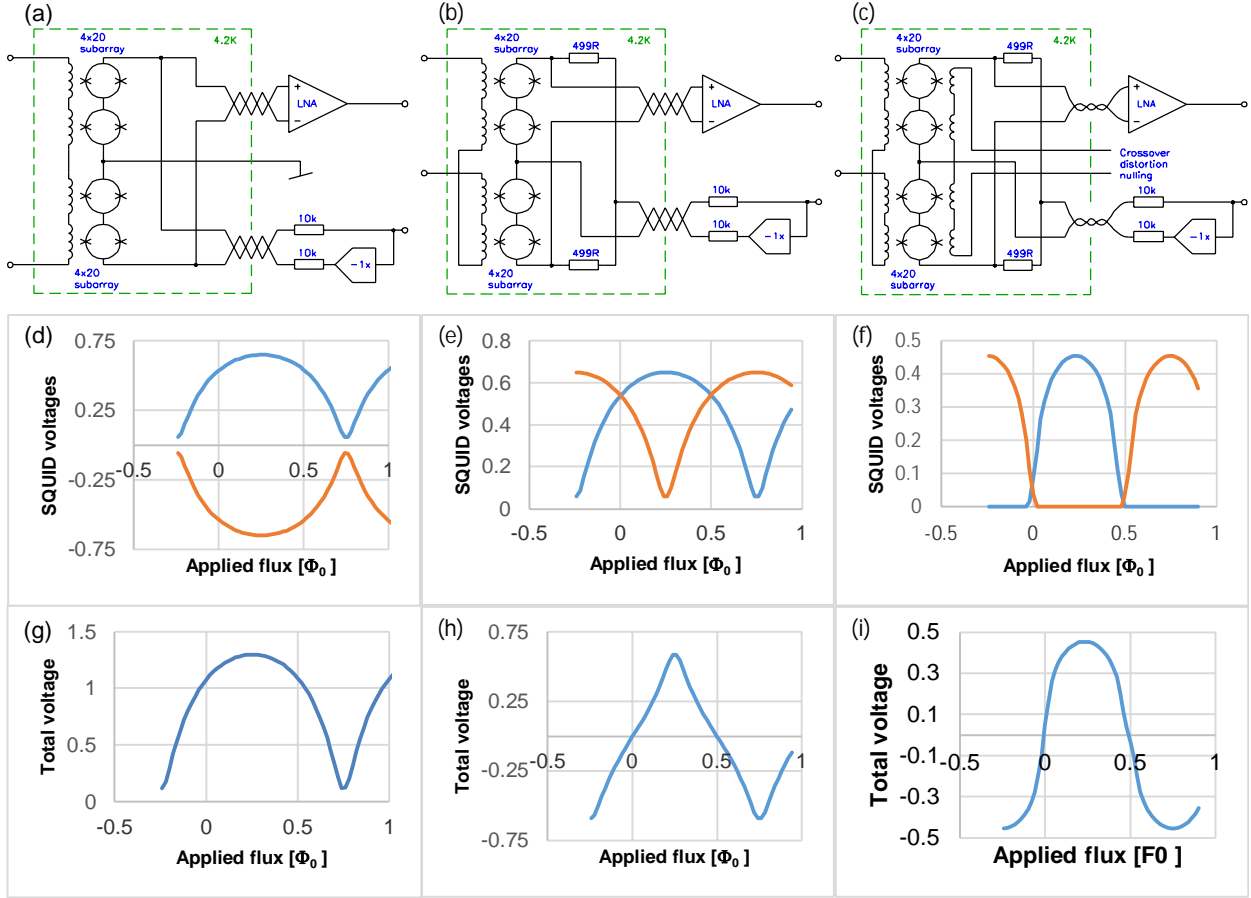

Figure 1: Schematics for the (a) class-A push-pull, (b) class-A push-push and (c) class AB/B push-push SQUID amplifier configuration. Panes (d), (e), (f): output voltages of the upper and lower SQUID sub-array in the three configurations, respectively, using numerically simulated flux characteristics of a $\beta_L = 1$, $\beta_C = 0.5$ SQUID as the response shape. Voltages are expressed in units of $R_S I_C$. Panes (g), (h), (i): the differential voltage seen by the LNA in the three configurations, respectively. A flux offset coil is used to bring the relevant parts of SQUID characteristics to the zero value of the applied flux, but not drawn in schematics (a), (b) and (c).

Numerically simulated flux characteristics of a dc SQUID with the dimensionless inductance $\beta_L = 1$ and McCumber-Stewart parameter (dimensionless capacitance) $\beta_C = 0.5$ are used to model the non-linearity. We have performed experimental tests with the 'N3'-type SQUID array, a version of the 'N2' [6] with the modified input coil polarity.

The class-B capable version of the push-push circuit is shown in figs. 1c, 1f and 1i. Here the bias currents of the sub-arrays are lowered until the SQUIDs remain in the zero voltage state over roughly half the flux range. Then current is directed to the crossover nulling (CN) coil, until the sub-array transitions to the finite voltage meet. The CN coil current can be adjusted such that both sub-arrays remain at a small but finite quiescent voltage even when the applied flux is zero, corresponding to the class-AB operation. Our 'N3'-type chip is equipped with such a CN coil. In a future circuit it is conceivable to guide the SQUID bias current through the CN coil so that a separate CN line will not be needed.

Here again, $\beta_C = 0.5$ numerical SQUID characteristics are used. Note however that a larger $\beta_C$ can be utilized without hysteresis when SQUID remains at a low bias current. We remark that the often-cited numerically calculated noise optimum for the dc SQUID [14] has been obtained with the nominally hysteretic $\beta_C > 1$ SQUID, operated at a low bias.

*2.1 Dissipation*

We have argued [4] based on the I-V characteristics of the dc SQUID that the static power dissipation is $P_D = 0.6\, R_S\, I_C^2$ in terms of the shunt resistance $R_S$ and one-junction critical current $I_C$, when a class-A SQUID is driving a LNA with close-to-matched input impedance $R_{IN} \approx 1.4 \times R_S$. Contrary to this, static dissipation of the class-B amplifier is strictly zero,

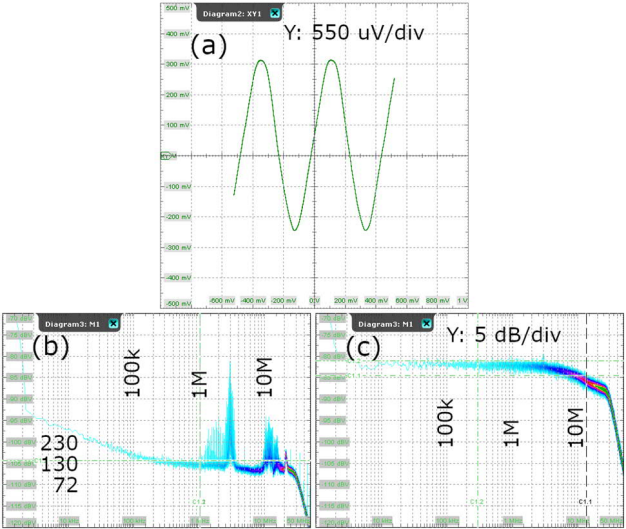

Figure 2: Experimental charcteristics from the Class-A push-push configuration of the 'N3' SQUID array. (a) Flux response. (b) Flux noise spectrum, in vertical units of $n\Phi_0/\text{Hz}^{1/2}$. (c) Frequency response obtained with pseudorandom flux noise excitation shows -3 dB bandwidth of 13 MHz.

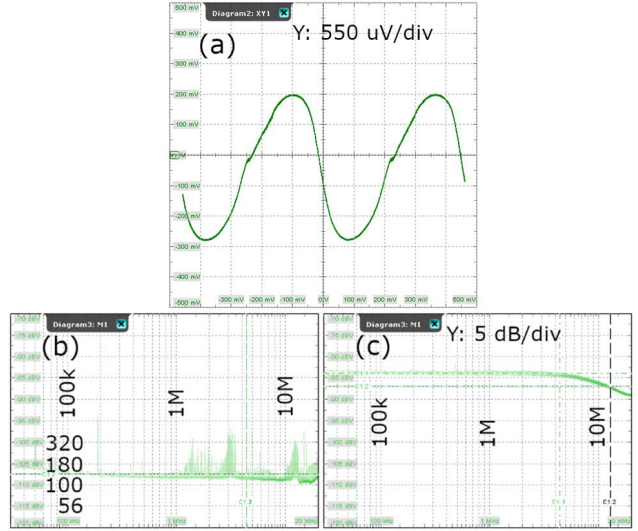

Figure 3: Experimental characteristics from the Class B push-push configuration of the 'N3' chip. (a) Flux response. (b) Flux noise spectrum, in vertical units of $n\Phi_0/\text{Hz}^{1/2}$. (c) Frequency response obtained with pseudorandom flux noise excitation shows -3 dB bandwidth of 12.5 MHz.

and can be made very small when operated in class-AB.

The low static dissipation is attractive eg. with X-ray microcalorimeter array readouts, in which the amplifier remains idle a significant fraction of time, waiting for X-ray events. In this regime, the obtainable dissipation in a millikelvin SQUID stage [4] may approach that of microwave multiplexers [15], quoted as several picowatts per pixel in Table 3 of [16]. We note that the SQUID heat budget allocation in the Frequency Domain Multiplexed class-A readout for the ATHENA mission [6] is 25 pW per pixel, and in practice the readout specification is reached on half this power [17]. Class-B readout could reduce the number significantly.

Numerical average over the characteristics fig. 1d, 1e and 1f with $\pm 0.125\ \Phi_0$ sinusoidal flux excitation, half the $\pm 0.25\ \Phi_0$ monotonous range, yield dynamic dissipation of $P_D = 0.5\, n\, R_S\, I_C^2$ for both class-A circuits, and $P_D = 0.14\, n\, R_S\, I_C^2$ for the class-B circuit. Here $n$ is the number of SQUIDs in the amplifier array.

## 3. Experiment

The constituent SQUIDs in the array are gradiometric with $L_{SQ}$ = 70 pH loop inductance, equipped with $0.8 \times 0.8\ \mu$m SWAPS Josephson junctions [18] at $J_C = 1.5 \times 10^7$ A/m$^2$ critical current density and $R_S = 17.5\ \Omega$ shunt resistors. The SQUIDs are equipped with a 3-turn input coil, providing $M^{-1}$ = 14 $\mu$A/$\Phi_0$ periodicity. The 'N3' chip contains two 20-series 4-parallel sub-arrays of these dc SQUIDs.

The LNA [19] uses a differential pair of discrete SiGe transistors as the input stage, shows $\Delta f >$ 40 MHz bandwidth, $u_N \approx 0.5$ nV/Hz$^{1/2}$ voltage noise, $i_N \approx 2$ pA/Hz$^{1/2}$ current noise and $G$ = 180 V/V gain. The dc offset voltage of the discrete transistor pair contributes to the observed non-zero LNA output voltage at the zero applied flux in experiments.

Fig. 2 depicts the flux response, noise spectrum and frequency response recorded for the 'N3'-type chip operated in the class-A configuration of the fig. 1b. The test was performed in liquid helium, using a dipstick wired with BeCu twisted pairs with ~20 $\Omega$ return circuit resistance. We estimate that the peak forest in the noise spectrum at a few MHz range is electromagnetic emission from the Rohde&Schwartz RTO2014 instrument which we have used in the experiment, as double checked by the HP89410A instrument. Origin of the $f >$ 10 MHz spectral peaks could not be verified because they reside above the HP89410A frequency range. Fig. 3 depicts the same characteristics when the SQUID array is operated in the class-B. We observe 0.12 $\mu\Phi_0$/Hz$^{1/2}$ and 0.15 $\mu\Phi_0$/Hz$^{1/2}$ flux noise in the two configurations, respectively, dominated by the LNA noise in both cases.

## 4. Conclusion

We have demonstrated practical operation of a class-AB push-push SQUID amplifier. Dissipation of the circuit can be estimated to be roughly 1/4 of a conventional class-A SQUID amplifier with comparable parameters when half-full range flux excitation is applied. In applications where duty cycle of the input excitation is low, such as readouts of single-photon detectors, the close-to-zero static dissipation suggests reduction in the average dissipated power by an order of magnitude or more. In this regime the class-B based multiplexer circuits may even be competitive against microwave multiplexers [15].

**Acknowledgement**
We thank Jan van der Kuur for discussions regarding the possible competitiveness of the class-B scheme against microwave multiplexers.

This work has been supported in part by the Academy of Finland through the Centre of Exellence for Quantum Technologies. The SQUID array has been fabricated under the Contract 4000123669/18/NL/BW by the European Space Agency.